\documentclass[twoside,11pt]{article}

\usepackage{jmlr2e}
\usepackage{gantt}
\usepackage{fancyvrb}
\usepackage{color}

\ShortHeadings{Anomaly Detection for malware identification using Hardware Performance Counters}{Alberto Garcia-Serrano}
\firstpageno{1}

\begin{document}

\title{Anomaly Detection for malware identification using Hardware Performance Counters}
% \date{March 8, 2015}
%\author{Alberto Garcia-Serrano\\ UOC}

 \maketitle

% Title
%\HRule \\[0.4cm]
%{ \huge \bfseries Anomaly Detection for malware identification using hardware performance counters \\[0.4cm] }
%\HRule \\[1.5cm]

% Author and supervisor
\noindent
\begin{minipage}{0.4\textwidth}
\begin{flushleft} \large
\emph{Author:}\\
Alberto \textsc{Garcia-Serrano}\\
\emph{email:}agarciaserra (AT) uoc.edu
\end{flushleft}
\end{minipage}%

\vskip 0.5in
\begin{abstract}
\emph{Computers are widely used today by most people. Internet based applications, like ecommerce or ebanking attracts criminals, who using sophisticated techniques, tries to introduce malware on the victim computer. But not only computer users are in risk, also smartphones or smartwatch users, smart cities, Internet of Things devices, etc.\\
Different techniques has been tested against malware. Currently, pattern matching is the default approach in antivirus software. Also, Machine Learning is successfully being used. Continuing this trend, in this article we propose an anomaly based method using the hardware performance counters (HPC) available in almost any modern computer architecture. Because anomaly detection is an unsupervised process, new malware and APTs can be detected even if they are unknown. 
}
\end{abstract}

{\bf Keywords:} Malware detection, Performance counters, Anomaly detection

\section{Introduction}
Hardware Performance Counters, HPC for short, are specialized registers implemented in modern processors. The aim of those register are to count different types of internal events in order to conduct performance analysis, algorithm optimization or software tuning. The available events and number of counters depends on the manufacturers. There are a lot of events, but only a few counters (two or four counters are common). Only one event can be assigned to a counter at once (although multiplexing techniques can be used to give the illusion of more than one event by counter). This means that only a few type of events can be measured while a single execution, so the events wanted have to be preselected.
As an example, some events that can be measured are: data cache missed, instruction cache missed, branch prediction faults, instruction executed, cycles, etc.
The observation of the counters can be useful for detecting some non-desirable states of a running software, like a malware attack. Malware usually uses a vulnerability in a program to modify the execution. Two of the most used techniques are Buffer Overflow \citep{citeulike:6345464} and ROP \citep{Prandini:2012:RP:2420631.2420866}. 
The buffer overflow techniques tries to bypass the limits of a memory buffer to illegally write in other zones of the program. The ROP (Return Oriented Programing) are used when the stack is not permitted to execute code, then little portions of code finished with the \emph{ret} instruction, also called gadgets, are used to build the shellcode from actual and yet existent program code. 
Both techniques modifies the natural flow of a program and leaves a footprint in the HPC measures.
It's possible, using Machine Learning Techniques to detect those anomalous behavior in the program execution. Two possible approaches can be used in Machine Learning: Supervised and unsupervised Learning. In Supervised Learning, a Classifier is trained with positive and negative data (i.e. non-malware programs and malware programs). Although a trained classifier is able to generalize, the classifier could have problems with unknown malware (like APTs). This approach is not very different from current antivirus software that needs to know the malware before be able to detect it.
In addition to the supervised learning, there exist Unsupervised Learning techniques in Machine Learning. In Unsupervised Learning it's not necessary to train the classifier since another strategy is taken. Instead of providing positive and negative examples, one, two or more features in the data are used to create groups with similar characteristics. This technique is named Clustering and let discover different classes of elements in a dataset. Also, there is other subfield in Machine Learning named Anomaly Detection \citep{Chandola:2009:ADS:1541880.1541882} whose techniques are able to find outliers in datasets. This is, data that are different from the rest because they deviate from the norm. Anomaly Detection is specially interesting because don't need previous training and could be able to find "strange" situations like those in which a malware exploit a vulnerability (e.g. while overflowing a buffer).

\section{State of the art}
Identifying malware attacks is a challenging problem. Different approaches have been used to address this task, but one of the most promising are those based on Machine Learning. One of the most used approach is detecting and analyzing automated activity based on the malware behavior. \cite{Rieck:2011:AAM:2011216.2011217} uses a sandbox to analyze the malware execution based on the system calls, generating a pattern expressed in q-grams and using a high-dimensional space to apply clustering methods. Similar approach has been taken by \cite{Eskin02ageometric} using a a multidimensional space of features and applying different algorithms (cluster-based estimation, K-nearest neighbor and One Class SVM) to successfully detect outliers in the feature space.
A comparison of Classification and Anomaly Detection to detect attacks has been done by \cite{Wressnegger:2013:CLN:2517312.2517316}. The experiments were done with web attacks, but they show that Anomaly Detection is a suitable approach. 
When using Classification or Anomaly Detection, the features taken into account are very important. Recently, some research groups has been working with performance signatures. \cite{Avritzer:2010:MSI:1712605.1712623} has analyzed the use of basic performance signatures like memory consumption, CPU usage or number of TCP connections to detect various types of attacks including buffer and stack overflow, SQL injection, DoS and MITM attacks. Other researchers have studied the problem of selecting the most appropriate features in attack detection, like \cite{DBLP:journals/corr/abs-0912-1014} that using the DARPA KDD CUP 99 dataset of network intrusions data taken by an IDS \citep{Tavallaee:2009:DAK:1736481.1736489} applied an ensemble approach based in information gain (from Shannon information theory). From an initial void subset of features, iteratively, the algorithm select those with more information gain and add it to the subset of features. Based on the same dataset that the previous researchers, in \cite{Ghali09featureselection} they prosed a new hybrid algorithm RSNNA (Rough Set Neural Network algorithm). The algorithm uses Rough Set theory in order to filter out superfluous and redundant information and a trained an artificial neural network to identify any kind of new treats.
With the recent addition of HPC to the modern processors there are new performance signatures available that can be used to detect malware. HPC are know to be accurate enough  \citep{conf/ispass/ZaparanuksJH09} and widely available and accesible with multiplatform libraries like PAPI \citep{Mucci99papi:a} or using standard tools like perf \citep{PerfExamples}.
\cite{Demme:2013:FOM:2508148.2485970} is probably one of the first attempts to use HPC for malware detection. They show that it's feasible to detect attacks measuring the HPC and using a classifier. They used different Machine Learning algorithms: Decision Tree, KNN, Random Forest and FANN. Decision Trees seems to be the most accurate algorithm detecting malware, still having a 10 percent of false positives. Recently, \cite{DBLP:journals/corr/TangSS14} used HPC but with an Anomaly Detection approach instead of Classification methods. Surprisingly, they suggest that any event anomalies manifested by malware code execution are not directly detectable, due to two key observations. 
\\(1) Most of the measurement distributions are very positively skewed, with many values clustered near zero. 
\\(2) Deviations, if any, from the baseline event characteristics due to the exploit code are not easily discerned. \\To address the problem they rely on rank-preserving power transform on the measurements to positively scale the values and successfully increase the detection percentage. Despite the results, there is still room to try new approaches using Anomaly Detection techniques and HPC that may show more promising and direct results.

\section{Research questions and objectives}
In the previous section, some detection methods have been presented that are being used in the problem of identifying malware attacks using Machine Learning techniques. We have reviewed both, Classification and Anomaly Detection methods. However, it's clear that a malware attack should be detected in an early stage (i.e. while exploiting a vulnerability) in order to mitigate it's undesirable effects. Anomaly Detection algorithms seems to be more sensible to unknown attacks and also, they don't need a previous training.
The question that this research tries to answer is if is it possible to use Anomaly Detection directly, without using any other statistical construction like the rank-preserving power transform used in \cite{DBLP:journals/corr/TangSS14}. In short, the objective of this research is to find a new method for detecting malware attacks directly based on Anomaly Detection techniques over captured HPC execution measures in a running program.

\section{Research methodology}
For the research process, using positivism as a philosophical paradigm, two methodologies are going to be used: Experiments and Design and Creation.
Experiments are used to find which techniques may be applied in this scenario and which of them best fit for this particular problem. This new knowledge will be used to establish a new method for the malware detection problem. Experiments will also be used to find which events in HPC contribute with more information for the Anomaly Detection algorithm.
Design and Creation will be used as research methodology to establish the new method (model and algorithm) for the malware detection problem.

\section{Experiments setup}
To measure HPC for a process while executing three alternatives have been evaluated:
\\(1) Instrument the software source code with a library like PAPI.
\\(2) Develop a kernel module that interrupt the process execution in a regular basis to measure HPC.
\\(3) Use the Linux \emph{perf} kernel utility.
\\Clearly, the option 3 is the easiest but there are some problems with the \emph{perf} utility that have to be solved before using it. With \emph{perf record} \citep{perfrecord} there is an option (-F) to profile the execution with a given frequency. After some test, we are unable to use this option successfully. An alternative is to use \emph{perf stat} \citep{perfstat}. In 2013, a patch was added to take measures of HPC at a given interval of time with the -I option \citep{perfpatch}. Since this is a valid way to take measures there is still a limitation of 100ms from measure to mesure by the IO latency. Since we are going to use a buffer to write the results to a file, we have recompiled the \emph{perf} utility to get down this limitation to 1ms.
\\
In order to set up a controlled environment for the experiments, we will use a simple but real program named nweb \citep{nweburl}, A tiny web server with modifications: the most important is that the web server has been modified to be single threaded in order to make it easier to measure the HPC.
The web software has a function named \emph{logger()} who is vulnerable to a stack overflow attack by the \emph{sprintf()} function in the next line

\begin{verbatim}
case LOG: (void)sprintf(logbuffer," INFO: %s:%s:%d",s1, s2,socket_fd); break;
\end{verbatim}

Malware usually use those type of programming errors to exploit the software. With nweb two types of common attacks are going to be used: Stack overflow and ROP (Return Oriented programming). In the next subsection the exploitation phase is described.

\subsection{Data acquisition}
The number and type of HPC events are CPU dependent, so for this experiment we used only a common subset shared by most CPUs. Table 1 contains the events we took into account for this study.

\begin{table}[h]
\begin{tabular}{l l l l}
cpu-cycles & instructions & cache-references & cache-misses\\ branches & branch-misses & bus-cycles & ref-cycles\\ L1-dcache-loads & L1-dcache-stores & L1-icache-loads & L1-icache-load-misses\\ LLC-loads & LLC-load-misses & LLC-stores & LLC-store-misses\\ dTLB-loads & dTLB-load-misses & dTLB-stores & dTLB-store-misses\\ iTLB-loads & iTLB-load-misses & branch-loads & branch-load-misses 
\end{tabular}
\caption{HPC Events}
\end{table}

For every event previously listed we execute some random page requests for a while to simulate a regular web browsing (the same requests pattern was used for every counter using a script in python) and finally, after the page requests, the attack was performed. The experiments were repeated at a HPC measuring frequency of 1ms, 10ms and 100ms.
Every execution was also repeated finishing with a classic stack overflow attack, a ROP attack and with a clean exit (without any attack). As a result, more than 400MB of text data has been recollected and analyzed.
As shown in the next example, the data generated at every execution contains the timestamp of the event, the event measure delta (increase from last measure) and the name of the event. 

\begin{verbatim}
# started on Sun Apr 19 01:23:16 2015

     0.001225993,1621,branch-load-misses
     0.002574349,5149,branch-load-misses
     0.003808515,5352,branch-load-misses
     0.005025360,5807,branch-load-misses
     ...
\end{verbatim}

\section{Data analysis}
Analyzing big amount of data looking for anomalies is an useful technique that can be applied to a large set of applications, including fraud detection, health improvement, marketing, biology, signal processing, etc. The goal of anomaly detection is to find one or more outliers. In the scope of this work, an outlier is an observation that deviates so much from other observations as to arouse suspicion that it was generated by a different mechanism other than the regular program execution flow.

\subsection{Anomaly detection approaches}

Cluster analysis is the task of grouping similar set of objects in a dataset. Those similar groups are called clusters, and are widely used in Machine Learning for unsupervised classification tasks. There are many different cluster models and many clustering algorithms for each model.    
Most used approaches are connectivity based, centroid based, distribution based and density based. Some of those clustering methods have been applied yet in \cite{DBLP:journals/corr/TangSS14} to the problem of detecting malware. \\
(1) Connectivity based (or hierarchical) clustering  uses the concept of distance to group the elements in sets of similar objects. The main idea is that nearby objects are more related that farther objects.\\  
(2) Centroid based clustering uses the concept of a central vector (usually fixed to a k number of vectors). The problem to solve is to find the position of the k central vectors which best separates the data space and optimize the data distribution. The most known algorithm of this type is k-means clustering.\\
(3) Distributed based clustering tries to cluster objects belonging to the same statistical distribution. One of the most known method is the Gaussian mixture model. This method suffers higher trend to overfitting than the others presented here.\\
(4) Density based clustering uses the concept of data density to group the objects. Most popular algorithms of this type are DBSCAN and OPTICS. There is also a local density approach called Local Outlier Factor (LOF). As far as we know, local density based methods have not been tested for malware detection with HPC.

\subsection{Local Outlier factor (LOF)}
As suggested by \cite{DBLP:journals/corr/TangSS14}, no significant deviations are detected with classical anomaly detection methods, but using statistical methods to amplify the small deviations, they improved the attack detection. We used a different approach: the local outlier factor or LOF \citep{Breunig:2000:LID:335191.335388}. Instead of finding global outliers like in the other density based clustering methods, this method tries to find anomalous data objects by measuring the local deviation of a given data object with respect to its neighbors.
The local outlier factor is based on the concept of a local density, where locality is given by k nearest neighbors. To estimate the density, the distance to those neighbors are taken into account. By comparing the local density of an object with the local densities of its neighbors, regions of similar density can be identified. Also, objects that have lower density than their neighbors can be detected. These are considered to be outliers.

\begin{figure}[h!]
  \centering
    \includegraphics[width=0.30\textwidth]{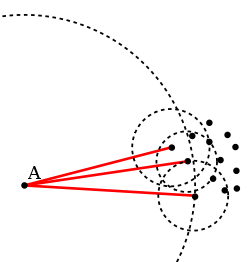}
    \caption{LOF: Object A has a much lower density than its neighbors}
\end{figure}

Let \mbox{k-distance}(A) be the distance of the object A to the k-th nearest neighbor. We define reachability distance as
\[\mbox{\[reachability-distance}_k(A,B)=\max\{\mbox{k-distance}(B), d(A,B)\}\]

Objects that belong to the k nearest neighbors of B are considered to be equally distant (see fig. 2).

\begin{figure}[h!]
  \centering
    \includegraphics[width=0.30\textwidth]{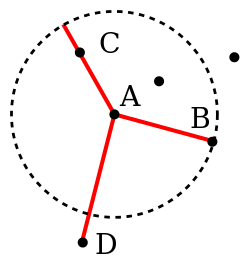}
    \caption{LOF: Objects B and C have the same reachability distance from A, taking k=3. D is not a k nearest neighbor}
\end{figure}

The local reachability density (ldr) of an object A is defined as the inverse of the average reachability distance of the object A from its neighbors.

\[\mbox{lrd}(A):=1/\left(\frac{\sum_{B\in N_k(A)}\mbox{reachability-distance}_k(A, B)}{|N_k(A)|}\right)\]

So we define the local outlier factor (LOF) for the object A (taking k neighbors) as the average local reachability density of the neighbors divided by the object's own local reachability density.

\[\mbox{LOF}_k(A):=\frac{\sum_{B\in N_k(A)}\frac{\mbox{lrd}(B)}{\mbox{lrd}(A)}}{|N_k(A)|} = \frac{\sum_{B\in N_k(A)}\mbox{lrd}(B)}{|N_k(A)|} / \mbox{lrd}(A)\]

This factor indicates a local density score comparing object A with its neighbors. If the factor takes the value 1, the object A has the same local density as it's neighbors. A larger value indicates that the object A is an outlier. The larger the value, the bigger the outlier. 

\section{Experiments results}
The HPC measuring frequency seems to be important to successfully find valid outliers. Surprisingly, reading the counters every 1ms or every 10ms don't increase the detection capacity, otherwise, the attack detection capacity falls when the measuring frequency is high. We found that reading HPC events every 100ms improve the efficiency of the LOF algorithm (and, obviously, makes the process faster and suitable for realtime analysis). We also found that using 5 neighbors points (k=5) for the LOF algorithm performs well for our needs. 
\\After the data analysis and comparison, six candidate counters were found: iTLB-load-misses, dTLB-loads, bus-cycles, LLC-store-misses, LLC-loads, LLC-load-misses.

\begin{figure}[h!]
  \centering
    \includegraphics[width=1\textwidth]{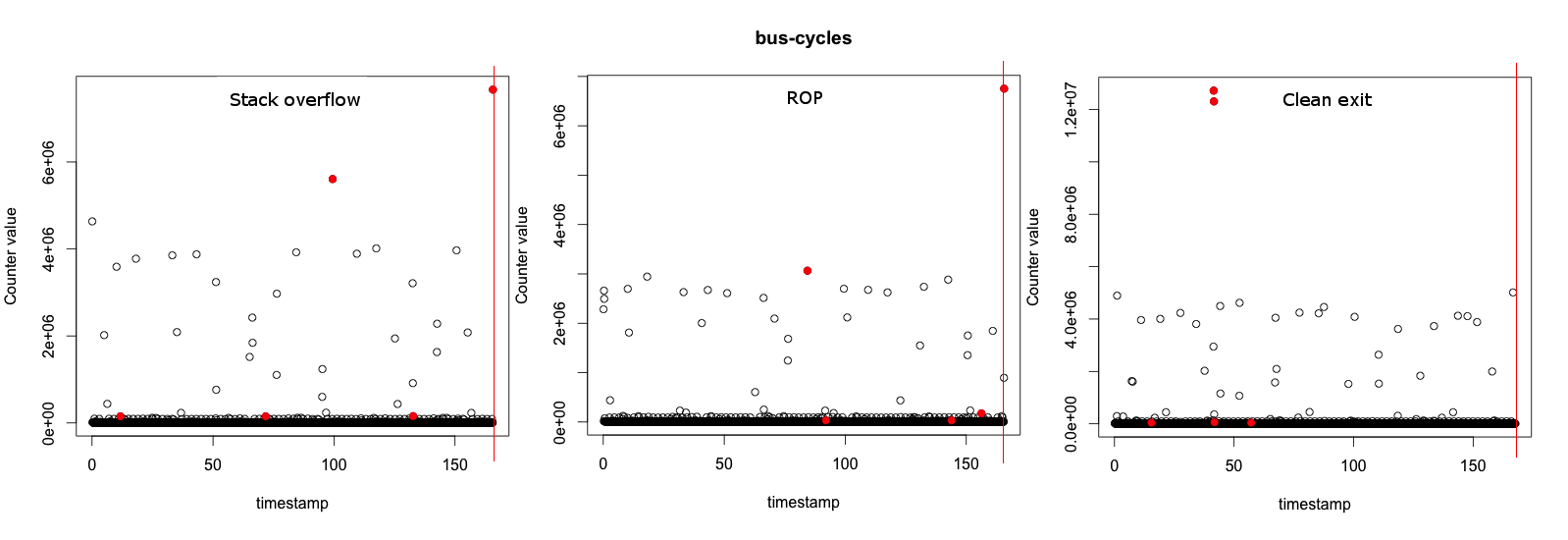}
    \caption{HPC cycles}
\end{figure}

\begin{figure}[h!]
  \centering
    \includegraphics[width=1\textwidth]{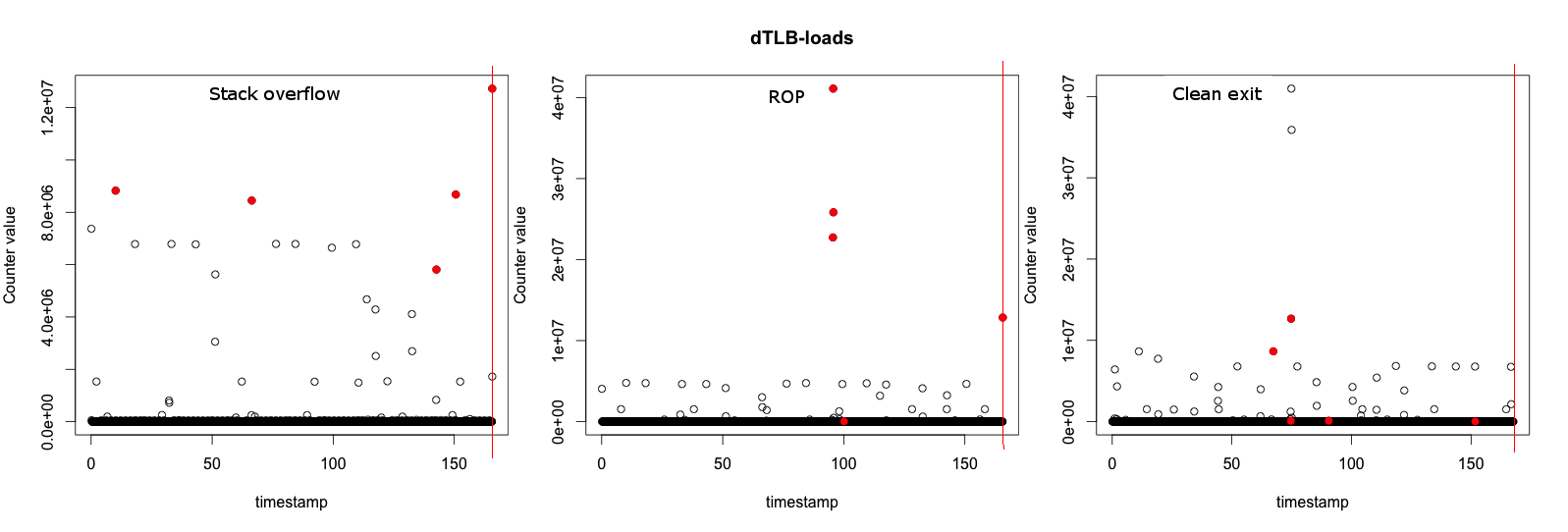}
    \caption{HPC dTLB-loads}
\end{figure}

\begin{figure}[h!]
  \centering
    \includegraphics[width=1\textwidth]{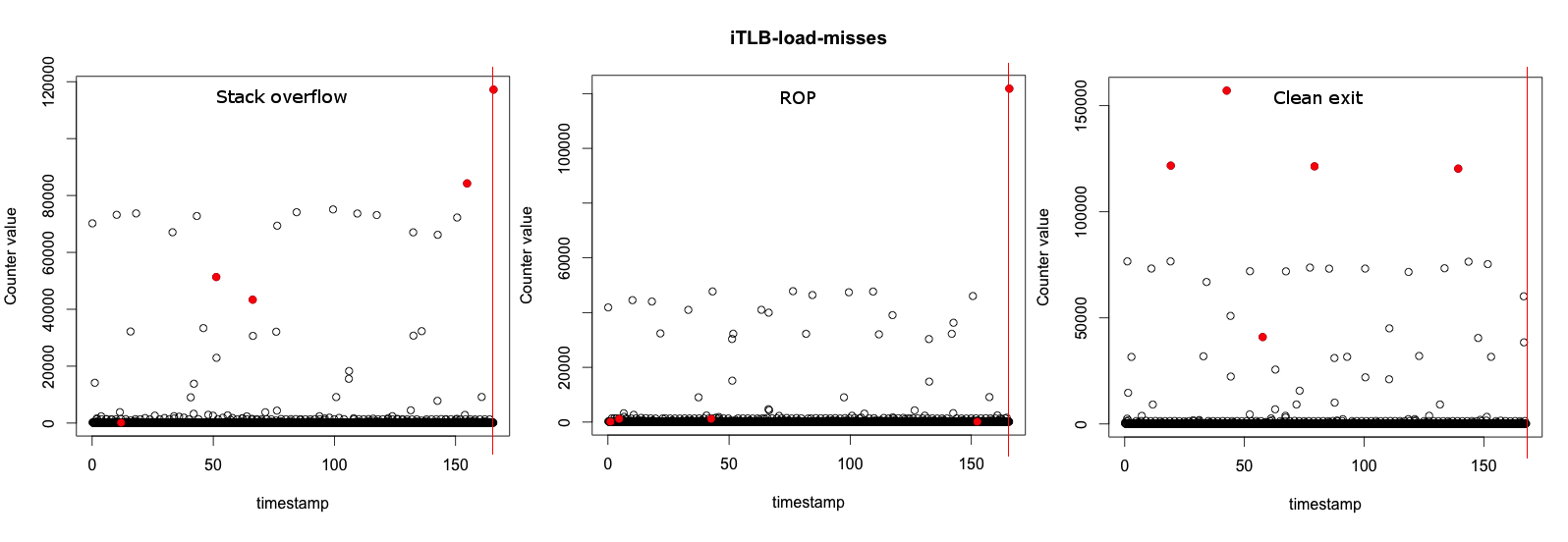}
    \caption{HPC iTLB-load-misses}
\end{figure}

\begin{figure}[h!]
  \centering
    \includegraphics[width=1\textwidth]{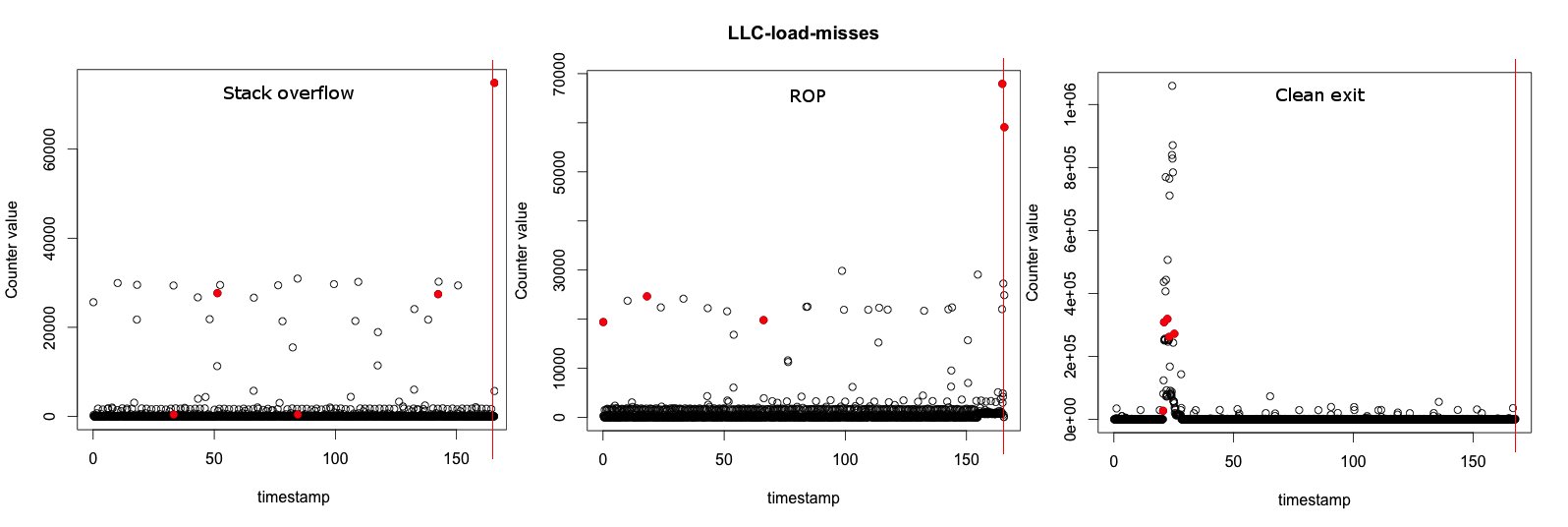}
    \caption{HPC LLC-load-misses}
\end{figure}

\begin{figure}[h!]
  \centering
    \includegraphics[width=1\textwidth]{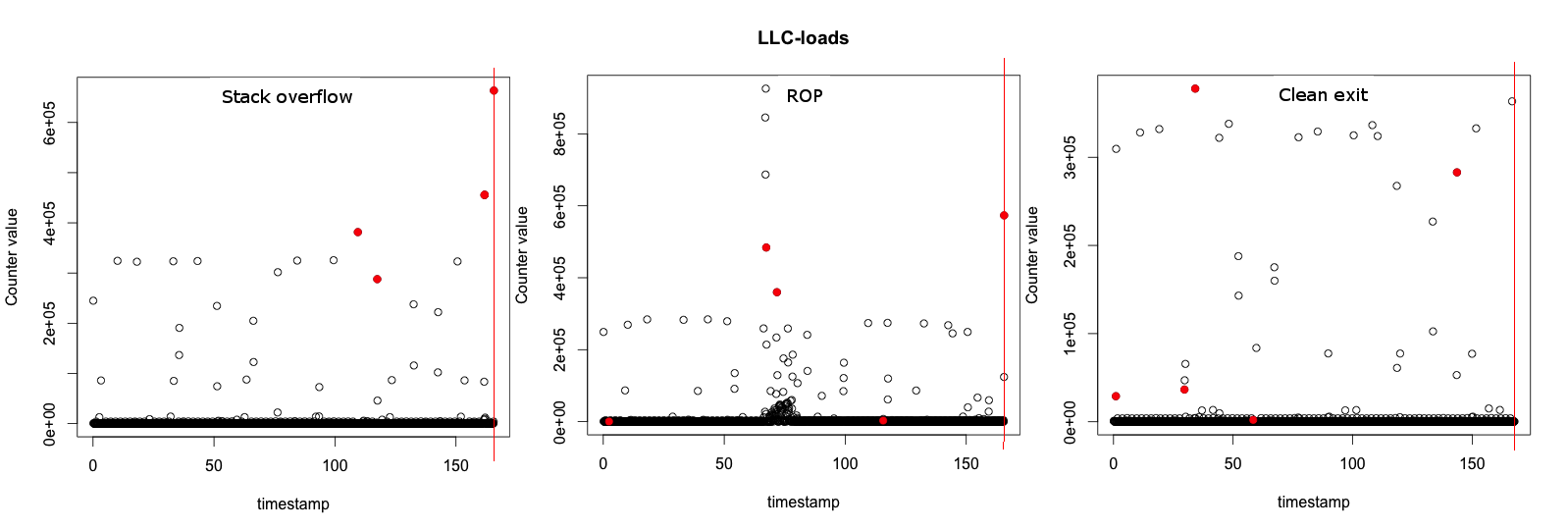}
    \caption{HPC LLC-loads}
\end{figure}

\begin{figure}[h!]
  \centering
    \includegraphics[width=1\textwidth]{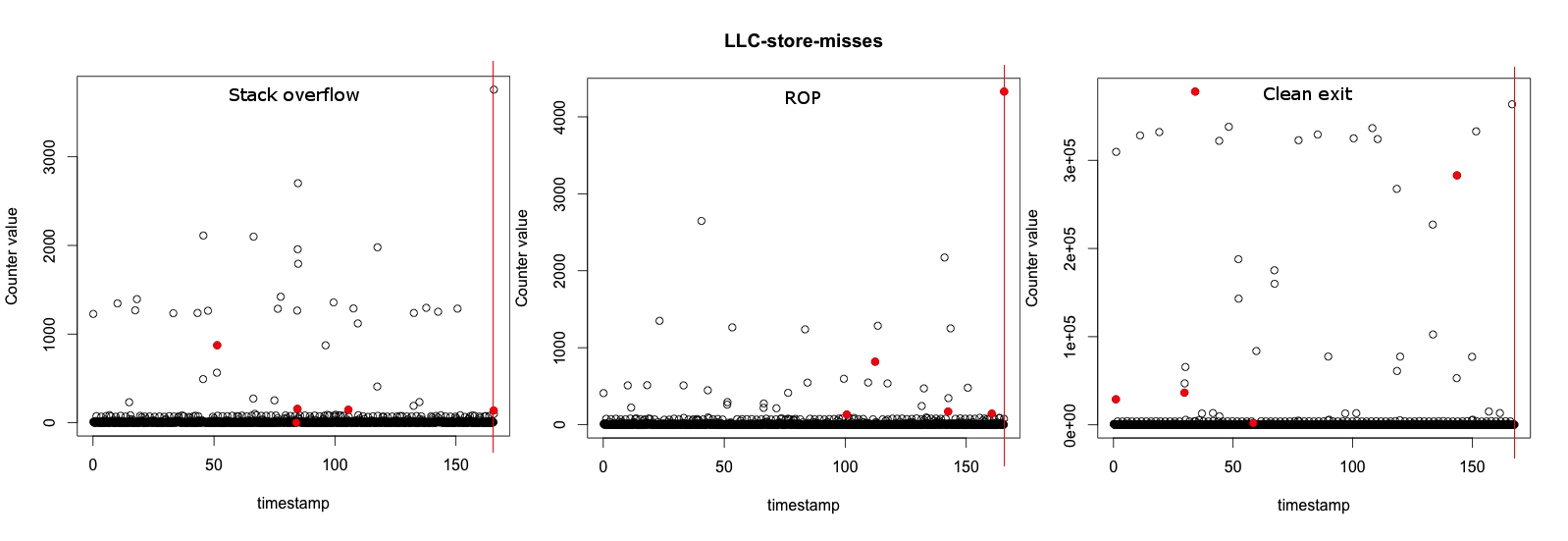}
    \caption{HPC LLC-store-misses}
\end{figure}

Figures 3 to 8 show the measures for the candidate events, where the 5 top outliers are represented by a filled red circle. A vertical red line marks the moment of the attack (or the program exit if a clean exit was performed).
The criterion to select those six counters as candidates was that outliers counter measures were detected while the attack, and also, no outlier were detected while the clean exit. In fact, those figures shows how an outlier measure is detected in the moment of the stack overflow and the ROP attack, but not at the moment of the clean exit, so it seem that it's feasible to detect an attack using those six HPC counters and the LOF algorithm.
\\As the result shows, the more useful counters for this task are related to cache operations (except the bus-cycles counter). iTLB refers to the instruction cache while dTLB is about data cache in the translation lookaside buffer (TLB), ie. a cache used when mapping virtual addresses to physical ones. It makes sense that the iTLB-load-misses counter is a good predictor because the exploit tries to change the natural program flow. 
\\LCC-* counters refers to the last level of the cache hierarchy (the largest but slower cache). Three of the candidate counters fall in this class of counters, so they proved to be better predictors of a program exploitation than L1 cache counters. 
\\Other HPC events like branch-misses seems to be useful, but trend to have a high level of false positives at the program exit phase. 

\section{Proposed method for malware detection}
Depending on the architecture, only a subset of the proposed HPCs can be used at once. A value of 2 or 4 may be typical. Let's n be the number of selected counters and k the number of neighbors used in the LOF calculation. We define the attack factor for the sequence time t as

\[F_{attack}(t-((k/2)+1))=\frac{\sum_{1}^{n}LOF_{k}(n)}{n}\]

We use the moment t-((k/2)+1)) because we need to evaluate the k neighbors in the time serie. This factor is the arithmetic mean of the LOF value for the n selected HPCs. Lets define a threshold value \(\delta > 1\). if \(F_{attack}(t-((k/2)+1)) > \delta\) there is a high provability of an ongoing attack.
A good value for \(\delta\) is 1.5 or above, but it's architecture dependent, so this value should be set by empirical observations. A value \(k=5\) worked fine in our experiments, but can be also tuned specifically for different architectures. 

\section{Conclusions and further work}
Previous research works has stated that it's possible to detect a malware attack using the HPC measures and Machine Learning. Those works used both supervised and unsupervised learning. The results demonstrated the feasibility of using HPC for this task, but no successful results were achieved without using additional statistical tools to amplify the small deviations produced by the malware. In this article we have used a different approach based on the LOF (Local Outlier Factor) density-based clustering method. We found that, with low false positive rate, it's possible to detect the attack with no additional statistical tools. We also found that not all counters are valid in the task of detecting malware. We found six of them to be valid counters (from those included in our study).\\ 
In conclusion, our experiments results shows that an ongoing stack attack can be detected almost in realtime using HPCs. Anyway, this is a preliminar work on the topic and further work should be done in order to tune up the proposed technique described in this article. Specifically \\
(1) Repeat the experiments with other different software programs (similar result may be expected).\\
(2) Repeat the experiments with different type of attacks. In this article we just tested our method with stack attacks because they are the more common attacks.\\
(3) Make a real implementation to test the method in real world conditions.\\
(4) Measure the real impact of false positives.

\vskip 0.2in

\bibliography{tfm}

\begin{thebibliography}{20}
\providecommand{\natexlab}[1]{#1}
\providecommand{\url}[1]{\texttt{#1}}
\expandafter\ifx\csname urlstyle\endcsname\relax
  \providecommand{\doi}[1]{doi: #1}\else
  \providecommand{\doi}{doi: \begingroup \urlstyle{rm}\Url}\fi

\bibitem[Avritzer et~al.(2010)Avritzer, Tanikella, James, Cole, and
  Weyuker]{Avritzer:2010:MSI:1712605.1712623}
Alberto Avritzer, Rajanikanth Tanikella, Kiran James, Robert~G. Cole, and
  Elaine Weyuker.
\newblock Monitoring for security intrusion using performance signatures.
\newblock In \emph{Proceedings of the First Joint WOSP/SIPEW International
  Conference on Performance Engineering}, WOSP/SIPEW '10, pages 93--104, New
  York, NY, USA, 2010. ACM.

\bibitem[Breunig et~al.(2000)Breunig, Kriegel, Ng, and
  Sander]{Breunig:2000:LID:335191.335388}
Markus~M. Breunig, Hans-Peter Kriegel, Raymond~T. Ng, and J\"{o}rg Sander.
\newblock Lof: Identifying density-based local outliers.
\newblock \emph{SIGMOD Rec.}, 29\penalty0 (2):\penalty0 93--104, May 2000.
\newblock ISSN 0163-5808.

\bibitem[Chandola et~al.(2009)Chandola, Banerjee, and
  Kumar]{Chandola:2009:ADS:1541880.1541882}
Varun Chandola, Arindam Banerjee, and Vipin Kumar.
\newblock Anomaly detection: A survey.
\newblock \emph{ACM Comput. Surv.}, 41\penalty0 (3):\penalty0 15:1--15:58, July
  2009.

\bibitem[Cowan et~al.(2000)Cowan, Wagle, Pu, Beattie, and
  Walpole]{citeulike:6345464}
C.~Cowan, F.~Wagle, Calton Pu, S.~Beattie, and J.~Walpole.
\newblock {Buffer overflows: attacks and defenses for the vulnerability of the
  decade}.
\newblock In \emph{DARPA Information Survivability Conference and Exposition,
  2000. DISCEX '00. Proceedings}, volume~2, 2000.

\bibitem[Demme et~al.(2013)Demme, Maycock, Schmitz, Tang, Waksman,
  Sethumadhavan, and Stolfo]{Demme:2013:FOM:2508148.2485970}
John Demme, Matthew Maycock, Jared Schmitz, Adrian Tang, Adam Waksman, Simha
  Sethumadhavan, and Salvatore Stolfo.
\newblock On the feasibility of online malware detection with performance
  counters.
\newblock \emph{SIGARCH Comput. Archit. News}, 41\penalty0 (3):\penalty0
  559--570, June 2013.

\bibitem[Eranian(2015)]{perfpatch}
Stephane Eranian.
\newblock perf patch.
\newblock \url{https://patchwork.kernel.org/patch/2004891/}, 2015.

\bibitem[Eskin et~al.(2002)Eskin, Arnold, Prerau, Portnoy, and
  Stolfo]{Eskin02ageometric}
Eleazar Eskin, Andrew Arnold, Michael Prerau, Leonid Portnoy, and Sal Stolfo.
\newblock A geometric framework for unsupervised anomaly detection: Detecting
  intrusions in unlabeled data.
\newblock In \emph{Applications of Data Mining in Computer Security}. Kluwer,
  2002.

\bibitem[Ghali(2009)]{Ghali09featureselection}
Neveen~I. Ghali.
\newblock Feature selection for effective anomaly-based intrusion detection.
\newblock In \emph{IJCSNS International Journal of Computer Science and Network
  Security, vol.9 no.3}, pages 285--289, 2009.

\bibitem[Gregg(2015{\natexlab{a}})]{perfrecord}
Brendan Gregg.
\newblock perf record.
\newblock \url{http://linux.die.net/man/1/perf-record}, 2015{\natexlab{a}}.

\bibitem[Gregg(2015{\natexlab{b}})]{perfstat}
Brendan Gregg.
\newblock perf stat.
\newblock \url{http://linux.die.net/man/1/perf-stat}, 2015{\natexlab{b}}.

\bibitem[Gregg(2015{\natexlab{c}})]{PerfExamples}
Brendan~D. Gregg.
\newblock perf examples.
\newblock \url{http://www.brendangregg.com/perf.html}, 2015{\natexlab{c}}.

\bibitem[Griffiths(2015)]{nweburl}
Nigel Griffiths.
\newblock nweb: a tiny, safe web server.
\newblock \url{http://www.ibm.com/developerworks/systems/library/es-nweb/},
  2015.

\bibitem[Mucci et~al.(1999)Mucci, Browne, Deane, and Ho]{Mucci99papi:a}
Philip~J. Mucci, Shirley Browne, Christine Deane, and George Ho.
\newblock Papi: A portable interface to hardware performance counters.
\newblock In \emph{In Proceedings of the Department of Defense HPCMP Users
  Group Conference}, pages 7--10, 1999.

\bibitem[Prandini and Ramilli(2012)]{Prandini:2012:RP:2420631.2420866}
Marco Prandini and Marco Ramilli.
\newblock Return-oriented programming.
\newblock \emph{IEEE Security and Privacy}, 10\penalty0 (6):\penalty0 84--87,
  November 2012.

\bibitem[Rieck et~al.(2011)Rieck, Trinius, Willems, and
  Holz]{Rieck:2011:AAM:2011216.2011217}
Konrad Rieck, Philipp Trinius, Carsten Willems, and Thorsten Holz.
\newblock Automatic analysis of malware behavior using machine learning.
\newblock \emph{J. Comput. Secur.}, 19\penalty0 (4):\penalty0 639--668,
  December 2011.

\bibitem[Singh and Silakari(2009)]{DBLP:journals/corr/abs-0912-1014}
Shailendra Singh and Sanjay Silakari.
\newblock An ensemble approach for feature selection of cyber attack dataset.
\newblock \emph{CoRR}, abs/0912.1014, 2009.

\bibitem[Tang et~al.(2014)Tang, Sethumadhavan, and
  Stolfo]{DBLP:journals/corr/TangSS14}
Adrian Tang, Simha Sethumadhavan, and Salvatore~J. Stolfo.
\newblock Unsupervised anomaly-based malware detection using hardware features.
\newblock \emph{CoRR}, abs/1403.1631, 2014.

\bibitem[Tavallaee et~al.(2009)Tavallaee, Bagheri, Lu, and
  Ghorbani]{Tavallaee:2009:DAK:1736481.1736489}
Mahbod Tavallaee, Ebrahim Bagheri, Wei Lu, and Ali~A. Ghorbani.
\newblock A detailed analysis of the kdd cup 99 data set.
\newblock In \emph{Proceedings of the Second IEEE International Conference on
  Computational Intelligence for Security and Defense Applications}, CISDA'09,
  pages 53--58, Piscataway, NJ, USA, 2009. IEEE Press.

\bibitem[Wressnegger et~al.(2013)Wressnegger, Schwenk, Arp, and
  Rieck]{Wressnegger:2013:CLN:2517312.2517316}
Christian Wressnegger, Guido Schwenk, Daniel Arp, and Konrad Rieck.
\newblock A close look on n-grams in intrusion detection: Anomaly detection vs.
  classification.
\newblock In \emph{Proceedings of the 2013 ACM Workshop on Artificial
  Intelligence and Security}, AISec '13, pages 67--76, New York, NY, USA, 2013.
  ACM.

\bibitem[Zaparanuks et~al.(2009)Zaparanuks, Jovic, and
  Hauswirth]{conf/ispass/ZaparanuksJH09}
Dmitrijs Zaparanuks, Milan Jovic, and Matthias Hauswirth.
\newblock Accuracy of performance counter measurements.
\newblock In \emph{ISPASS}, pages 23--32. IEEE, 2009.

\end{thebibliography}

\end{document}